\documentclass[%
reprint,
 amsmath,amssymb,
 aps,
 prl,
 showpacs,
]{revtex4-1}

\usepackage{graphicx}
\usepackage{dcolumn}
\usepackage{bm}

\begin{document}


\title{A method to calibrate the absolute energy scale
 of air showers with ultra-high energy photons}

\author{Piotr Homola}
 \affiliation{University of Siegen, Siegen, Germany}
 \affiliation{H.~Niewodnicza\'nski Institute of Nuclear Physics PAN,  Krak\'ow, Poland}

\author{Markus Risse}%
 \affiliation{University of Siegen, Siegen, Germany
}%

\date{\today}

\begin{abstract}
Calibrating the absolute energy scale of air showers initiated by
ultra-high energy cosmic rays is an important experimental issue.
Currently, the corresponding systematic uncertainty amounts to
14-21\% using the fluorescence technique.
Here we describe a new, independent method which can be applied if ultra-high energy
photons are observed. While such photon-initiated showers have not yet
been identified, the capabilities of present and future cosmic-ray
detectors may allow their discovery.
The method makes use of the geomagnetic conversion of UHE photons
(preshower effect), which significantly affects the subsequent
longitudinal shower development. The conversion probability depends
on photon energy and can be calculated accurately by QED.
The comparison of the observed fraction of converted photon
events to the expected one allows the determination of the absolute
energy scale of the observed photon air showers and, thus, an
energy calibration of the air shower experiment.
We provide details of the method and estimate the accuracy that
can be reached as a function of the number of observed photon showers.
Already a very small number of UHE photons may help to test and fix
the absolute energy scale.
\end{abstract}

\pacs{14.70.Bh, 95.55.Vj, 95.85.Ry, 96.50.sb}
\maketitle


Measuring the cosmic-ray flux and spectral features gives
important clues on the origin of ultra-high energy (UHE) cosmic rays.
For this, a sufficiently good energy reconstruction 
of the air showers initiated by UHE cosmic rays is needed which
is an experimental challenge.
Using the fluorescence technique, ideally in coincidence with a
detector array, giant air shower observatories
achieve a systematic uncertainty of 14$-$21\% in determining
the absolute energy scale
\cite{auger-energyscale-icrc2013,hires-energy-scale,ta-energyscale-icrc2013}.
This is based on a piece-by-piece calibration of all relevant components
in the reconstruction chain which includes the
fluorescence yield in air, the transparency of the atmosphere,
optical properties and the electronic response of the detector,
and a correction for the invisible shower energy.

Comparing the present flux spectra from different experiments to each other,
one finds a fairly good agreement when allowing for shifts in energy
within the quoted systematic uncertainties \cite{cern2013-spectrum-wg}.
This shows that {\it relative} energies can be determined quite well
by shower experiments.
However, for getting better clues on the origin of UHE cosmic rays,
a direct comparison of data to theoretical predictions in terms
of an {\it absolute} energy scale is needed.
This comparison is limited by the experimental systematic uncertainty
quoted above.

As an example, a currently open question is whether 
cosmic rays at highest energy are (mostly) protons and whether the
observed flux suppression above 4$\times$$10^{19}$~eV
is due to photo-pion production with the CMB during
propagation (GZK effect \cite{gzk1,*gzk2}).
Based on this ansatz, very detailed theoretical predictions about
the energy spectrum above $\sim$$10^{18}$~eV were made 
\cite{berezinsky88,*berezinsky06},
including the absolute energies of prominent features in the spectrum
like a ``dip'' (due to pair production) or a ``cutoff energy $E_{1/2}$''
(due to photo-pion production).
A confirmation of these predictions to high precision would
constitute significant evidence for the existence of UHE protons
and for their interaction with the CMB. This, in turn,
would have profound implications for future research in terms of
cosmic-ray astronomy at UHE and searches for GZK neutrinos and photons.
In fact, a dip-like feature (ankle) and a suppression at UHE are
observed \cite{hires-energy-scale,hires-gzk,ta-gzk,auger-gzk-prl-2008,auger-icrc2013-spectrum}. However, alternative, completely different scenarios
can also explain the observed spectra within the given 
uncertainties \cite{aloisio11,allard2012}.
Reducing the experimental systematic uncertainty of the
absolute energy scale could significantly help to test the different
scenarios and clarify the situation.

Here we present and study the potential of a new,
completely independent method that could allow a direct
end-to-end calibration of the absolute energy scale at highest energy
above $\sim$2$\times10^{19}$~eV.
It makes use of the quite sharp threshold behaviour of pair
production by UHE photons
in the magnetic field of the Earth (geomagnetic conversion, or preshower).
This is a well-known process of standard physics that can be calculated
accurately by QED.

For the new method to work, one needs
(1) a sufficiently good determination of the relative shower energy,
of the shower direction and a sufficiently good separation between
three classes of events (converted photons, unconverted photons, non-photons)
and
(2) the observation of UHE photon events.
While the techniques for (1) are available as also discussed below,
an observation of UHE photons is still lacking
(see \cite{kalmykov_2013,*fomin_2013}, though).
However, current experiments have a growing sensitivity for the
detection of UHE photons 
(e.g.\ \cite{risse-homola-rev-07,auger-sd-photon-limits,icrc2011-auger-photons,cern2013-multimessenger-wg})
both due to increasing data sets and due to an improved
measurement of air showers by upgrading the detectors.
And, as we will see, already a very small number of UHE photon events
may help to test and fix the absolute energy scale.

The total probability $P_{\rm conv}$ of the UHE photon conversion
is closely related to the primary photon energy $E$. 
It can be concluded from e.g.\ \cite{erber} that
$P_{\rm conv}(E)$ is a strictly increasing 
function of $E$ for terrestrial geomagnetic conditions 
up to the maximum cosmic-ray energies observed by now.
Particularly, $P_{\rm conv}(E)$ shows a quite sharp threshold behaviour,
with $P_{\rm conv}$ increasing from 10\% to 90\% within a factor $\sim$2
in energy.
The key idea here is that one can determine in turn
the absolute photon energy
$E = E(P_{\rm conv})$ provided $P_{\rm conv}$ can be measured
to sufficient precision.

\begin{figure}
\begin{center}
\includegraphics[width=0.4\textwidth]{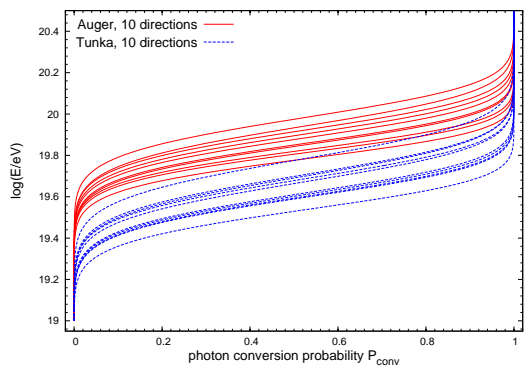}
\end{center}
\caption{Examples of the relation $E(P_{\rm conv})$ for random arrival directions at 
the locations of the Pierre Auger Observatory (weak local geomagnetic field)
and Tunka experiment (extremely strong local geomagnetic field). 
}
\label{e_vs_pconv_examples}
\end{figure}

Fig.~\ref{e_vs_pconv_examples} shows examples of this inverse relation $E(P_{\rm conv})$
for random arrival directions 
at two different observatory locations: the Pierre Auger Observatory \cite{auger-prototype-04} in Argentina, 
and the Tunka experiment \cite{Berezhnev:2012ys} in Russia. 
The calculations were performed
with the program PRESHOWER \cite{cpc1}. 
The local geomagnetic field is significantly different at the two sites: 
$\sim$0.23~G at Auger and $\sim$0.58~G at Tunka.
This is reflected by the energy shift in lines in Fig.~\ref{e_vs_pconv_examples}.
On the other hand, it can be seen that the relations plotted in Fig.~\ref{e_vs_pconv_examples}
have similar shapes in terms of the slope steepness 
$s \equiv {\rm d}E / {\rm d}P_{\rm conv}$.
The slope is important for the proposed method, as it
affects the precision of finding $E$: for a
given uncertainty $\Delta P_{\rm conv}$ in measuring $P_{\rm conv}$,
a small value of $s$ results in a small 
uncertainty of $E$, since $\Delta E \simeq s \Delta P_{\rm conv}$. 
For instance, assuming an uncertainty
$\Delta P_{\rm conv}=0.1$ one can estimate the relative energy uncertainty
for different values of $P_{\rm conv}$: 
$\Delta E(P_{\rm conv}=0.5)/E\simeq0.05$,
$\Delta E(P_{\rm conv} = 0.1$ or $0.9)/E \simeq 0.2$ and
$\Delta E(P_{\rm conv}=0.01$ or $0.99)/E \simeq 1.0$.
It turns out that the range $0.1 \leq P_{\rm conv} \leq 0.9$ 
where the slope $s$ is sufficiently small,
is most sensitive for an accurate energy determination.



We now show how to determine $P_{\rm conv}$ and $\Delta P_{\rm conv}$
from photon observations so that $E$ and $\Delta E$ can be concluded.
As an illustration of the method,
let us first consider the artificial case of observing
$n \gg 1$ photon air showers of same primary energy and arrival direction.
Out of these, $k$ are observed to be initiated by preshowers (primary photon converted),
$0 \ll k \ll n$. 
The probability of observing $k$ converted photon events out of $n$ photon showers
is given by a binomial 
probability distribution
with the maximum at $P_{\rm conv}=k/n$.
In turn, the observed ratio $k/n$ is the best estimate for $P_{\rm conv}$,
and $E(P_{\rm conv})$ can be concluded.
The uncertainty $\Delta P_{\rm conv}$ can be found by checking the cumulative binomial distributions for different 
values of $P_{\rm conv}$ and finding the distributions for which the observed $k$ can be excluded at a specified confidence level.
For example, considering 10 simulated photon events arriving at the Pierre Auger Observatory
from geographical South at a zenith angle of 13$^\circ$, out of which 3 are classified as converted, one gets
an absolute energy of
$E(n=10, k=3) \simeq 8.4^{+11\%}_{-8\%} \times 10^{19}$~eV.


In a realistic scenario, the photon events (in total perhaps just a few)
have different energies and arrive from different directions.
In this case a measurement of a single value of $P_{\rm conv}$ for a certain direction
is not possible anymore.
Nevertheless, the energy calibration
based on a measurement of just one number describing a ``global'' photon conversion rate
is feasible as long as the shower experiment provides a sufficiently good
measurement of {\it relative} shower energies.
All shower energies are affected by the same factor when adjusting
the absolute energy scale, and it is this factor that needs
to be determined.

Consider $n$ photon events numbered by index $i=1\,...\,n$ out of which $k$ were identified as converted. 
Based on current reconstruction methods
(using e.g.\ signal strength in fluorescence or ground detectors),
an initial estimate of energies $E_{\rm ini}(i)$ is obtained.
While the relative energies are determined sufficiently good,
the values $E_{\rm ini}(i)$ might differ from the true primary energies,
but all of them differ by the same factor $f_{\rm opt}$.
The purpose of the following method is to determine $f_{\rm opt}$.

We define the set $\{C\}$ 
of photon event classes:
$C(i)=0$ for an unconverted photon
and $C(i)=1$ for a converted photon. We have $\sum\limits_i C(i)=k$.
For given conversion probabilities $P_{\rm conv}(i)$ the probability of observing a specific set $\{C\}$ is given by:
\begin{equation}
Q(C)=\prod\limits_{i_{\rm conv}=1}^k P_{\rm conv}(i_{\rm conv})\prod\limits_{i_{\rm unconv}=k+1}^{n} (1-P_{\rm conv}(i_{\rm unconv})) 
\label{eq-ptot}
\end{equation}
where $i_{\rm conv}$ and $i_{\rm unconv}$ number the converted and unconverted events, respectively.

To find $f_{\rm opt}$ and its uncertainty and, thus, to get the absolute energy
of the photon events, the following procedure is adopted
(``bootstrapping'' approach):\\
(1)
The initially estimated photon energies $\{E_{\rm ini}(i)\}$ are multiplied 
by a factor $f_j$ to generate a set of shifted energies $\{E_{\rm shift}(i,j)\}$: $E_{\rm shift}(i,j)=f_j \cdot E_{\rm ini}(i)$. 
Having the set of shifted energies we use the relation $P_{\rm conv}(E)$
to compute the corresponding values of conversion probabilities $\{P_{\rm conv}(i,j)\}$
and $Q_j$ with Eq.~\ref{eq-ptot}. \\
(2) We repeat step 1 with different factors $f_{j}$ (e.g.\ $f_{j} \simeq 0.7\,...\,1.3$
given the present experimental uncertainties of 14$-$21\%).
The optimum shift $f_{\rm opt}$ is the one
which maximizes $Q_j(\{P_{\rm conv}(i,j)\})$ given by Eq.~\ref{eq-ptot}. 
This energy shift fits best the observation (i.e.\ $\{C\}$).\\ 
(3)
We proceed and estimate $\Delta f_{\rm opt}$.
The found $f_{\rm opt}$ determines a set of conversion probabilities $\{P_{\rm conv}^{\rm opt}(i)\}$. Each of the 
probabilities $P_{\rm conv}^{\rm opt}(i)$ determines the expected class of the observed photon event: converted
(probability of occurrence $P_{\rm conv}(i)$)
or unconverted (probability of occurrence: $1-P_{\rm conv}(i)$). 
We generate $n_r$ random sets $\{C_r\}$ of ``conversion flags''
(i.e.\ photon class ``converted'' or ``unconverted''),
with $r=1\,...\,n_r$ numbering the sets.\\
(4)
We repeat steps 1 and 2 for each set $\{C_r\}$ and get the 
distribution of the corresponding optimum energy shifts $\{f_{\rm opt}^r\}$.
The width of this distribution is used to 
estimate the uncertainty (resolution) $\Delta f_{\rm opt}$.

We note that a two-sided confidence
interval is obtained only if both $i_{\rm conv}>0$ and $i_{\rm unconv}>0$ which is
the case we focus on in this paper.
For $i_{\rm conv}=0$ or $i_{\rm unconv}=0$ (i.e.\ all photons either converted
or unconverted), a one-sided confidence interval results.
This can still serve to place interesting limits to the energy scale
e.g.\ when just two or three unconverted photons at the lower end of
the preshower energy range are observed. We leave this special case
of just one event class in the data set,
which is increasingly unlikely for growing $n$,
for a further study.



We now evaluate the performance of the method (resolution and bias)
for data sets of size $n = 3\,...\,100$ adopting realistic experimental
conditions (see e.g.\ \cite{auger-energyscale-icrc2013,auger-icrc2013-photons-kuempel,auger-2010-xmax,
hires-2010-proton-composition,ta-tsunesada-2013icrc-composoition-summary,deltanmu-yakutsk-2010,deltanmu-amiga-2008}).
We take a resolution of 8\% in energy and $1^\circ$ in direction.
Furthermore, we account for uncertainties in classifying the 
events (converted photons, unconverted photons, non-photons)
by means of the shower observables $X_{\rm max}$
(depth of shower maximum) and $N_\mu$ (number of ground muons)
and take a resolution of $\Delta X_{\rm max}=20$~g~cm$^{-2}$ and
$\Delta N_\mu/N_\mu=0.2$.
We note that alternative observables might be used as well, as for
instance signal risetime and curvature of shower
front~\cite{auger-sd-photon-limits}.

Based on simulations using CONEX~\cite{conex} with 
QGSJet~II.04~\cite{qgsjet}, we determined the relevant misclassification
rates of
(i) non-photons misclassified as photons,
(ii) photons misclassified as non-photons,
(iii) converted (unconverted) photons misclassified as unconverted
    (converted) photons.
In essence, a simple but effective classification can be done
by using $N_\mu$ to separate photons from non-photons (typical
difference of factor $\sim$6 in $N_\mu$)
and by using $X_{\rm max}$ to separate converted from unconverted
photons (typical difference of $\sim$200~g~cm$^{-2}$ in $X_{\rm max}$).
From this approach, which could be improved to further
optimize the classification, we obtain the following rates:
(i) Based on $N_\mu$, the fraction of protons misclassified as photons
is $\sim$$10^{-4}$ (even smaller for heavier nuclei). Thus, the contamination
of the photon sample by hadrons can be kept sufficiently small for
cosmic-ray photon fractions down to the percent level or below.
(ii) For the same $N_\mu$ cut, the fraction of photons misclassified
as non-photons is below 10\% which implies a minor loss of photon event
statistics. As the misclassification rates are somewhat different
for unconverted photons ($\sim$6\%) and converted photons ($\sim$2\%),
this could lead to a slight overestimation of the energy scale
($\sim$5\% for $n = 5$ and $\sim$1\% for $n \ge 10$).
However, this bias can be corrected for, or reduced with more
sophisticated cuts.
(iii) Based on $X_{\rm max}$, the misclassification rate between
converted and unconverted photons is below 10\%. The specific cut values
can be adjusted in such a way that the expected number of misclassified
converted photons equals that of misclassified unconverted photons.
In this way, no bias is introduced.

\begin{figure}
\vspace{-0.2cm}
\begin{center}
\includegraphics[width=0.4\textwidth]{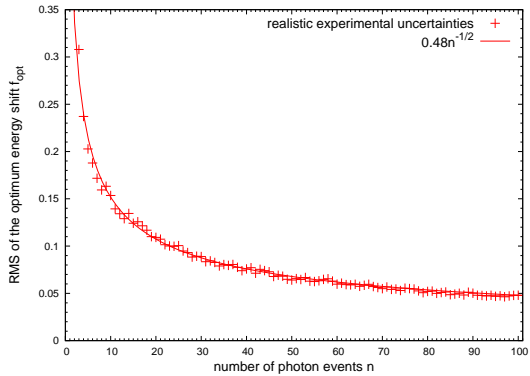}
\end{center}
\vspace{-0.2cm}
\caption{Resolution of the energy calibration method as a function of the
number $n$ of photon events, taking experimental uncertainties into account (see text). The resolution is well represented
by the function 0.48/$\sqrt n$ (solid line).
}
\label{eshift_rms_vs_n_varcomb}
\end{figure}

Taking these experimental uncertainties into account, the 
performance of the method (steps 1$-$4) to calibrate the
absolute energy of UHE
photons has been determined by Monte Carlo simulations.
The resulting resolution is shown in Fig.~\ref{eshift_rms_vs_n_varcomb}
as a function of the size of the data set. The simulations
were performed for the conditions of the Pierre Auger Observatory
for photons above 4$\times$$10^{19}$~eV with spectral index $-$4.2 (according to Ref.~\cite{auger-hybrid-spectrum-settimo})
and shower zenith
angle $<$$80^\circ$ (\cite{auger-icrc2013-spectrum}).
As including events with $P_{\rm conv}$ values close to 0 or 1 would vary the
total probability (Eq.~\ref{eq-ptot}) only slightly within a
wide range of energy shifts and would not improve the accuracy of
determining the optimum shift $f_{\rm opt}$, only events with
$0.1 \leq P_{\rm conv}(f_j=1.0) \leq 0.9$ are counted.
As mentioned earlier, we restrict our analysis to data sets which
contain both classes of photon events (relevant only for
very small $n$).
As can be seen from Fig.~\ref{eshift_rms_vs_n_varcomb},
the resolution improves  $\propto (1/\sqrt{n})$.
It is $\sim$20\% for $n = 5$, $\sim$14\% for $n = 10$, and
$<$10\% for $n > 20$.
Already with few events,
an accuracy comparable to current approaches is reached.

We checked that the residual bias of the method, after correcting
for the small effect from different misclassifications rates
of converted
and unconverted photons as non-photons discussed before in (ii),
is below $\sim$1\% for $n \geq 10$.
For smaller $n$, an overestimation of the energy scale of a few percent
appears (up to $\sim$10\% for $n = 3$).  This, however, can be understood
due to the requirement of having both classes of photons in the sample,
which leads to a small bias against unconverted photons (more numerous
at lower energy). Again, a correction could be applied to
account for and reduce this bias. In any case, the bias is well below the
resolution which just depends on sample size. Thus, the method is not
limited by systematics
but only by statistics.

The accuracy of the method varies little when changing the high energy interaction model
(e.g.\ EPOS-LHC~\cite{epos-lhc1,*epos-lhc2} instead QGSJet~II.04), the spectral index
(e.g.\ from $-$4.2 to $-$3) or increasing the detector resolutions
in energy (e.g.\ from 8\% to 15\%), arrival direction (e.g.\ from $1^\circ$ to $2^\circ$),
$X_{\rm max}$ (e.g.\ from 20~g~cm$^{-2}$ to 30~g~cm$^{-2}$) and in $N_\mu$
(e.g. changing $\Delta N_\mu/N_\mu$ from 0.2 to 0.3).

After calibrating the absolute energy scale with primary photons,
the energy reconstruction of primary hadrons needs to account for the
missing energy. The corresponding correction introduces an additional
systematic uncertainty of $\sim$1.5\% at
$10^{20}$~eV~\cite{auger-energyscale-icrc2013}. 



In summary,
we presented a new method to determine the {\it absolute} energy scale
of air showers initiated by UHE cosmic rays.
It is an end-to-end calibration working directly in the 
interesting regime of highest energy.
The method exploits the
one-to-one relation between the probability for pair production by UHE photons
in the geomagnetic field and the primary energy of these photons. 
The method is statistics-limited.
Already with a small number of events,
the accuracy of the method is comparable or superior to
current approaches.
Present and planned giant air shower experiments offer an unprecedented
sensitivity to detect UHE photons.
There is no guarantee that indeed UHE photons will be observed
in future. But if so, they may become the ``standard candle''
for calibrating air shower energies.


\begin{acknowledgments}

This work was supported 
in Germany by the Helmholtz Alliance for Astroparticle Physics and by the BMBF Verbundforschung Astroteilchenphysik. The authors
are grateful to their colleagues from the Pierre Auger Collaboration.

\end{acknowledgments}





\bibliography{references_ph140122}

\end{document}